\documentclass[acmsmall,screen]{acmart}
\AtBeginDocument{%
  }

\usepackage{multirow}
\usepackage{tabularx}
\usepackage{hyperref}
\usepackage{subcaption}
\usepackage{tcolorbox}

\begin{document}

% \title{Unit Test Generation With Large Language Models: A Systematic Literature Review}
% \title{Large Language Models for Unit Test Generation: Achievements, Challenges, and the Road Ahead}
\title{Large Language Models for Unit Test Generation: Achievements, Challenges, and Opportunities}

% \affiliation{%

\author{Bei Chu}
\affiliation{%
  \institution{State Key Laboratory for Novel Software Technology, Nanjing University}
  \city{Nanjing}
  \country{China}}
\orcid{0009-0000-0480-1495}
\email{beichu@smail.nju.edu.cn}

\author{Yang Feng}
\authornote{Yang Feng and Kui Liu are the corresponding authors.}
\affiliation{%
  \institution{State Key Laboratory for Novel Software Technology, Nanjing University}
  \city{Nanjing}
  \country{China}}
\orcid{0000-0002-7477-3642}
\email{fengyang@nju.edu.cn}

\author{Kui Liu}
\authornotemark[1]
\affiliation{%
  \institution{Software Engineering Application Technology Lab, Huawei}
  \city{Hangzhou}
  \country{China}}
\orcid{0000-0003-0145-615X}
\email{kui.liu@huawei.com}

\author{Zhaoqiang Guo}
\affiliation{%
  \institution{College of Computer Science and Technology, Zhejiang University}
  \city{Hangzhou}
  \country{China}}
\orcid{0000-0001-8971-5755}
\email{guozhaoqiang@zju.edu.cn}

\author{Yichi Zhang}
\affiliation{%
  \institution{School of Computer Science, Peking University}
  \city{Beijing}
  \country{China}}
\orcid{0009-0006-6536-0885}
\email{zhangyichi@stu.pku.edu.cn}

\author{Hange Shi}
\affiliation{%
  \institution{State Key Laboratory for Novel Software Technology, Nanjing University}
  \city{Nanjing}
  \country{China}}
\orcid{0009-0000-3092-0906}
\email{hangeshi@smail.nju.edu.cn}

\author{Zifan Nan}
\affiliation{%
  \institution{Software Engineering Application Technology Lab, Huawei}
  \city{Hangzhou}
  \country{China}}
\orcid{0000-0001-8568-236X}
\email{nanzifan@huawei.com}

\author{Baowen Xu}
\affiliation{%
  \institution{State Key Laboratory for Novel Software Technology, Nanjing University}
  \city{Nanjing}
  \country{China}}
\orcid{0000-0001-7743-1296}
\email{bwxu@nju.edu.cn}

% \renewcommand{\shortauthors}{Chu et al.}

\begin{abstract}

Unit tests help detect faults and prevent regressions, but writing them manually is costly. Automated unit test generation involves constructing inputs that exercise the code and oracles that express expected behavior. Large language models (LLMs) can use code and natural-language context to construct tests, but generated tests may fail to compile or execute, contain invalid assertions, or have limited ability to detect faults. We conducted a systematic literature review and mapping study of 178 publications with eligible full-text reports available by 31 July 2026. We organize the literature using a lifecycle-oriented analytical framework that covers test construction and quality assurance. Across the reviewed studies, context enrichment addresses missing project information, while iterative validation and repair use compiler and execution feedback to revise generated tests. These mechanisms help make tests executable, although their contributions to oracle correctness and fault detection vary across tasks and evaluation settings. Meanwhile, benchmark validity, test maintainability, and industrial reliability remain open concerns. Finally, we discuss opportunities in evaluation, context and specification use, hybrid techniques, and agentic workflows.

\end{abstract}






\maketitle

\section{Introduction}
\label{sec:intro}

Software testing is a fundamental engineering practice for assuring software quality and mitigating release risks~\cite{myers2011art,daka2014survey}. Unit testing is a test level that focuses on the smallest independently testable units of a system, such as a function, method, or class~\cite{beck2022test}. The tests may be designed using structural information, externally observable behavior, or both. A well-designed suite of unit tests can detect logic errors and boundary-case defects early in development, prevent regressions as the software evolves~\cite{yoo2012regression}, and support agile practices such as test-driven development~\cite{beck2022test}. Manually writing comprehensive, high-quality unit tests nevertheless remains costly and labor-intensive, as prior surveys report that it can consume more than 15\% of development time~\cite{daka2014survey,runeson2006survey}.

To address this challenge, researchers have long studied automated test generation (ATG), including search-based software testing (SBST)~\cite{fraser2011evosuite, lukasczyk2022pynguin, herlim2022citrus} and symbolic or concolic execution~\cite{takashima2021syrust, garg2013feedback, yoshida2017klover, rho2024taming}. These techniques have proven effective at systematically exploring program structures. However, their effectiveness depends on the objectives, constraints, and environment models available. Domain-specific formats, complex object states, and external interactions may require additional specifications, models, or manual engineering~\cite{fraser2011evosuite, harman2015achievements, mcminn2004search, zhang2020survey, baldoni2018survey, fraser2014large}, often limiting their ability to generate realistic test cases.

Large language models (LLMs) provide a complementary way to use learned code and natural-language patterns during test construction~\cite{chen2021evaluating, tufano2020unit}. LLM-based systems have generated inputs and fixtures~\cite{li2025based,wang2026fixturize}, test prefixes~\cite{pan2025aster,zhang2025citywalk}, and mock implementations for external dependencies~\cite{pan2025aster,gorla2025cubetesterai}. Project tests and documentation can supply examples of API use and descriptions of expected behavior to guide this generation~\cite{zhang2026reference,hossain2025doc2oracll}. These uses motivate the focus of this survey and require compilation, execution, and oracle validation of the generated tests.

Existing surveys cover LLM applications across software engineering~\cite{fan2023large,zhang2023survey}, while surveys of automated test generation compare established generation paradigms~\cite{anand2013orchestrated,chen2026survey}. More focused reviews summarize the use of LLMs in unit testing~\cite{zapkus2024unit}. For example, Zhang et al.~\cite{zhang2025large} examine testing tasks, model utilization, and supporting techniques. Our lifecycle analysis asks how project context, generated candidates, and validation feedback interact during test generation and how established testing techniques support this process. Section~\ref{sec:related} compares these surveys in detail.

Through a systematic analysis of 178 publications, this paper develops a lifecycle-oriented taxonomy of LLM-based unit test generation. The synthesis focuses on publications in which unit-test generation is a primary or co-primary objective. The taxonomy organizes the reviewed methods into test-construction and quality-assurance categories without assuming that every system follows a fixed sequence. Test-construction approaches usually draw on project context, retrieval, static analysis, slicing, and feedback, whereas quality-assurance approaches often use compilers, test runners, search and program-analysis tools, and coverage or mutation feedback. Within these categories, we compare the information provided to a generator, the test artifact it constructs, and the checks used to accept or revise that artifact. Compilation and execution are used to check whether a test builds and runs. However, oracle correctness, fault detection, and industrial applicability remain open challenges.

To structure our analysis, we address the following research questions (RQs):

\begin{itemize}
    \item \textbf{RQ1:} How do LLMs support the core phases of test construction, including context analysis, input generation, and oracle construction?
    \item \textbf{RQ2:} What mechanisms are employed to validate, repair, and optimize the quality of LLM-generated tests?
    \item \textbf{RQ3:} What are the primary challenges in current LLM-based unit-test generation, and what opportunities do they present for future research?
\end{itemize}

\section{Background}
\label{sec:background}

To situate LLM-based methods, we first review established automated unit test generation approaches and their documented limitations, and then summarize the capabilities and risks relevant to LLM-based test generation.

\subsection{Traditional automated test generation}
\label{sec:utgen}

Automated unit test generation can be analyzed around two central artifact components: a \textit{test prefix}, which establishes objects, inputs, and calls needed to exercise the code under test (CUT), and a \textit{test oracle}, which encodes expected behavior, commonly through assertions~\cite{barr2014oracle, bertolino2007software, myers2011art}. Fixtures, setup and teardown code, mocks, and environment configuration may support these components. For a test of \texttt{max(a,b)}, the prefix supplies numerical arguments and invokes the function, while the oracle checks the returned value. A function that makes an asynchronous network call can additionally require a mock that reproduces the session's behavior~\cite{wang2026fixturize}. The prefix--oracle distinction nevertheless provides a useful analytical basis for examining how generation methods construct executable scenarios and how they infer or obtain expected outcomes. Among established automated unit-test generation methods, this survey uses two prominent paradigms to frame the comparison with LLM-based approaches:

\paragraph{\textbf{Search-Based Software Testing (SBST)}}
SBST frames test case generation as a search-based optimization problem~\cite{harman2012search, mcminn2004search}. It employs metaheuristic algorithms, such as genetic algorithms, to evolve a set of test cases that maximize a fitness function, which is typically a code coverage metric like branch coverage~\cite{mcminn2004search, harman2015achievements}. Tools such as EvoSuite~\cite{fraser2011evosuite, fraser2014large} for Java and Pynguin~\cite{lukasczyk2022pynguin} for Python have demonstrated the effectiveness of this approach in exploring program structures. However, coverage-oriented fitness functions do not directly encode program intent~\cite{harman2015achievements}. SBST may also require many trials when a target depends on domain-specific formats, such as a valid IP address. Constructing complex object states can also be difficult because a test prefix may require a precise sequence of method calls within a large search space~\cite{herlim2022citrus, fraser2014large}. External interactions pose a related challenge when their behavior cannot be represented adequately in the fitness function~\cite{fraser2014large}.

\paragraph{\textbf{Symbolic and Concolic Execution}}
Symbolic execution provides a systematic approach based on formal methods~\cite{baldoni2018survey,zhang2020survey}. It executes a program with symbolic variables, collects path constraints, and uses a solver to generate inputs for selected paths. Its principal scalability challenge is path explosion: the number of feasible paths can grow rapidly with program size, making exhaustive exploration computationally infeasible. Concolic execution interleaves concrete and symbolic execution to guide the exploration of new paths, but does not fully eliminate path explosion~\cite{sen2005cute,godefroid2005dart}. Tools like KLEE and Coyote C++ illustrate symbolic and concolic approaches for bug finding and high structural coverage~\cite{zhang2020survey,rho2024taming}.

\subsection{LLMs for code generation}
\label{sec:llm4code}

SBST optimizes explicit objectives, whereas symbolic execution solves encoded path constraints. Project examples and specifications offer complementary guidance when testing requires domain-specific inputs, complex object construction, or models of external services~\cite{harman2015achievements,baldoni2018survey,fraser2014large}. LLM-based generation offers one such complement. Large-scale pre-training exposes LLMs to statistical patterns in code and aligned code--test data~\cite{chen2024survey,rao2023cat}. These patterns can guide them to generate semantically plausible inputs or construct complex objects. LLMs can also adapt their output based on examples provided in a prompt through in-context learning without parameter updates~\cite{austin2021program,ni2024casmodatest}. Project context can also supply input constraints, mocking information, and dependency details needed to construct and execute a test in the target environment. For example, LISP derives input partitions for library APIs~\cite{li2025based}, while ASTER and CubeTesterAI use project and library information to generate setup or mock code~\cite{pan2025aster,gorla2025cubetesterai}.

However, LLM generation is probabilistic, and generated tests may fail to compile, reference non-existent APIs, or violate the intended behavior~\cite{schafer2023empirical, yuan2024evaluating}. Incorrect assertions can also arise from incomplete context, ambiguous specifications, model limitations, or a lack of execution feedback in the generation setup~\cite{li2024large,huang2024measuring}. These limitations motivate the validation and repair mechanisms discussed in Section~\ref{sec:rq2}.

\section{Research Methodology}
\label{sec:researchmeth}

\begin{figure}[t]
    \centering
    \includegraphics[width=\linewidth,keepaspectratio]{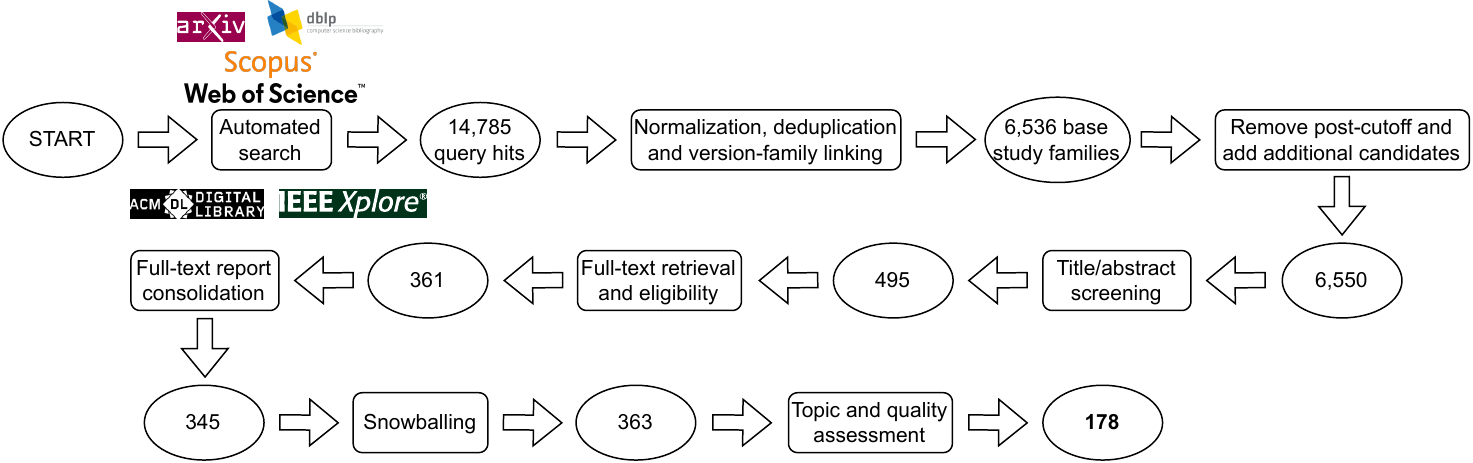}
    \caption{Overview of the paper collection process.}
    \label{fig:paper_collect}
\vspace{-0.3cm}
\end{figure}

\subsection{Paper collection methodology}
\label{sec:papercoll}

Fig.~\ref{fig:paper_collect} illustrates the systematic literature collection and screening process adopted for this survey. Final synthesis eligibility was content-based: an eligible LLM or code model had to generate a newly constructed unit-level test artifact, and unit-test generation had to be a primary or co-primary research focus. We required an accessible English full text with sufficient technical or empirical content. Publication venue and status were recorded as study characteristics and did not determine inclusion. The review cutoff was fixed at 31 July 2026, and reports of the same underlying work were linked into one study family.

The four core electronic sources were Scopus, Web of Science Core Collection, the ACM Digital Library, and IEEE Xplore. arXiv and DBLP were used for separately logged supplementary discovery searches. All six searches were last run on 18 August 2026. Before cross-query and cross-source deduplication, Scopus contributed 3,141 query hits, Web of Science 1,882, the ACM Digital Library 1,032, IEEE Xplore 4,300, arXiv 3,571, and DBLP 859.

The electronic searches used one broad Boolean query family that required at least one model term and at least one unit-test or test-artifact term. Table~\ref{tab:search_blocks} reports the exact conceptual blocks. Scopus applied their conjunction to \texttt{TITLE-ABS-KEY}, and Web of Science applied it to the \texttt{TS} topic field. The ACM Digital Library and IEEE Xplore used source-specific translations or partitions to accommodate their interface and query-length constraints; these adaptations retained the same search concepts, and all partitioned results were unioned before deduplication. arXiv used a corresponding partitioned query for supplementary discovery, while DBLP used documented title/metadata variants because its interface cannot express the complete nested Boolean query. No lower date limit or filter on venue, document type, language, or access was imposed during retrieval. The review cutoff and the requirement for an English full text were applied after retrieval during record management and eligibility assessment. The \href{https://anonymous.4open.science/r/csur_utgen-3F52}{anonymous replication package} provides exact queries, execution dates, counts, and logs.

\begin{table*}[t]
\centering
\caption{Exact conceptual blocks used in the electronic search. Entries within each block are joined by OR, and the two blocks by AND. The final entry retains its internal AND.}
\label{tab:search_blocks}
\footnotesize
\begin{tabularx}{\textwidth}{@{}>{\raggedright\arraybackslash}X@{}}
\toprule
\textbf{Model terms.} \mbox{``large language model*''}, LLM, LLMs, \mbox{``foundation model*''}, \mbox{``code language model*''}, \mbox{``large code model*''}, \mbox{``language model* for code''}, \mbox{``pretrained language model*''}, \mbox{``pre-trained language model*''}, \mbox{``pretrained code model*''}, \mbox{``pre-trained code model*''}, \mbox{``generative AI''}, \mbox{``generative artificial intelligence''}, \mbox{``AI coding assistant*''}, \mbox{``code assistant*''}, transformer*, \mbox{``deep learning''}, \mbox{``DL-based''}, \mbox{``DL based''}, \mbox{``GitHub Copilot''}, Copilot, ChatGPT, GPT, Codex, CodeT5, PLBART, BART, CodeBERT \\
\addlinespace[3pt]
\textbf{Unit-test artifact terms.} \mbox{``unit test*''}, \mbox{``unit testing''}, \mbox{``test generation''}, \mbox{``generating tests''}, \mbox{``generate tests''}, \mbox{``test case generation''}, \mbox{``test cases generation''}, \mbox{``test suite*''}, \mbox{``test code''}, \mbox{``assertion generation''}, \mbox{``assert statement*''}, \mbox{``test oracle*''}, \mbox{``test prefix*''}, \mbox{``test input*''}, \mbox{``test completion''}, \newline ((\mbox{``test case*''} OR \mbox{``test driver*''} OR \mbox{``test harness*''}) AND (generat* OR synthes*)) \\
\bottomrule
\end{tabularx}
\vspace{-0.3cm}
\end{table*}

We used a predefined venue list only to audit the coverage of the electronic searches. The journal list comprised TSE, TOSEM, EMSE, JSS, IST, STVR, Automated Software Engineering, and Software Quality Journal. The conference and workshop list comprised ICSE, FSE/ESEC-FSE, ASE, ISSTA, ICST, AST, SANER, ICSME, SSBSE, QRS, and LLM4Code. The list covers general software engineering, software testing and analysis, maintenance and evolution, search-based software engineering, and research on LLMs for code. It was neither presented as exhaustive nor used as a ranking of venue quality. Studies found by this audit were screened under the same content-based criteria as electronically retrieved records, irrespective of venue or publication status.

We then applied bidirectional snowballing to a provisional seed set retained after an initial check of topic relevance and the critical quality criteria, using reference lists for backward searching and citation-index records with publisher or repository checks for forward searching. Newly included studies seeded further iterations until a complete iteration yielded no new eligible study.

At both the title/abstract and initial full-text stages, two authors independently assessed each record before reconciliation. Disagreements were resolved by consensus against the recorded criteria. Candidates reaching final assessment through electronic searches, coverage audits, and snowballing were subjected to the same final topic and quality criteria.

\subsection{Collection results}
\label{sec:collres}

Fig.~\ref{fig:paper_collect} depicts the resulting record flow. The source-level counts entering normalization totaled 14,785 query hits. Normalization, deduplication, and preliminary version linking produced 6,536 unique candidate records. Removing 47 post-cutoff records and adding 61 candidates from the coverage audits produced 6,550 records for title/abstract screening, of which 495 advanced to retrieval. Four unavailable and two non-English reports were removed. Of the remaining 489 full texts, 128 failed the initial eligibility assessment, leaving 361 provisionally relevant reports. Consolidating 16 additional versions produced 345 search-derived publication candidates. Bidirectional snowballing identified 19 reports, one of which was linked to an existing publication, yielding 18 additional candidates. No further eligible publication was found in the next iteration. Applying the final topic and quality criteria uniformly to the resulting 363 candidates excluded 153 for scope and 32 for quality, leaving 178 publications for synthesis.

Unless stated otherwise, figures describing the corpus and statements about method distributions use the 178 included publications as their denominator. Studies outside the synthesis corpus may be cited to position maintenance, refactoring, industrial, or broader testing settings, but they are not pooled into prevalence estimates or treated as support for general effects. Statistics reported from an individual publication retain that publication's experimental denominator. Each count for search, screening, report consolidation, or author verification refers to the relevant stage population. For example, the earlier ChatTester report~\cite{yuan2023more} and its formal publication~\cite{yuan2024evaluating} counted as one publication. The formal version represented the work in the synthesis, with the earlier report retained in the report-level audit trail.

\subsection{Data extraction, quality assessment, and analysis}

\begin{table}[t]
\centering
\caption{Quality-appraisal criteria used in study assessment.}
\label{tab:quality_appraisal}
\small
\begin{tabularx}{\textwidth}{@{}>{\raggedright\arraybackslash}p{0.8cm}>{\raggedright\arraybackslash}X@{}}
\toprule
\textbf{ID} & \textbf{Quality criterion} \\
\midrule
QC1 & Sufficiently complete scholarly account of the claimed contribution. \\
QC2 & Clear reporting of model identity, access mode, configuration, and role. \\
QC3 & Description of datasets or systems under test, sampling, exclusions, and evaluation units. \\
QC4 & Explicit generated test artifacts and focal callable or unit boundaries. \\
QC5 & Reproducible description of the generation or repair pipeline. \\
QC6 & Evaluation design and baselines appropriate to the claims. \\
QC7 & Addresses compilation or execution, oracle validity, flaky or invalid tests, or equivalent checks. \\
QC8 & Complete, traceable metrics, denominators, and results. \\
\midrule
QC9 & Robustness checks, repeated runs, statistics, or justified alternatives reported. \\
QC10 & Code, prompts, data, tests, or other research artifacts available or sufficiently documented. \\
QC11 & Proportionate discussion of threats to validity and limitations. \\
\bottomrule
\end{tabularx}
\vspace{-0.3cm}
\end{table}

The author team completed one 11-item quality appraisal for each candidate retained after the initial topic check. After the initial quality gate, the team completed one primary structured extraction against RQ1--RQ3 for each retained candidate. Each criterion in Table~\ref{tab:quality_appraisal} received 2 points if fully met with traceable evidence, 1 if partially met or incompletely reported, and 0 if not met or affected by a serious concern. All scored candidates had 11 applicable criteria, giving a maximum score of 22. Inclusion required unit-test generation to be a primary or co-primary objective, QC1--QC8 each to score 2, a total score of at least 18/22, and a focal output consisting of a newly constructed complete unit test, prefix, fixture, input, assertion, or oracle. QC10 contributed to the total score but was not a separate inclusion requirement. Standalone modification of existing tests or suites and other uses that did not construct a focal unit-level artifact failed the focal topic criterion.

Because extraction was not independently duplicated for the full provisional corpus, we conducted a second-pass audit of 60 studies. The audit covered all records flagged for additional review in the earlier quality appraisal and a stratified sample across analytical layers and uses in the synthesis. It confirmed the recorded extraction, quality, taxonomy, and execution-validation classifications. To calibrate the taxonomy, both authors independently classified 20 of the audited studies without seeing the existing classifications. This produced no change to the analytical classifications.

We also recomputed the three principal findings and the coarse method-cluster distribution for all 178 publications. To check sensitivity to publication status and reported execution validation, we repeated the analysis for 131 publications with formal reports by the cutoff and 83 reporting explicit compilation, test execution, coverage, mutation, or bug-detection results. All six RQ1 analytical groups were represented in the full set. Seven of nine RQ2 groups met the focal synthesis criteria. The two omitted groups concerned standalone repair of existing broken tests and augmentation of existing suites. Eight of nine RQ3 groups remained. The omitted group concerned testers acting as verifiers in broader code/test co-evolution. These topics are retained only for contextual comparison.

The formal subset retained the RQ1 and RQ2 coverage patterns but one fewer RQ3 group because the relevant study using an adversarial mutation benchmark had no formal report by the cutoff. The execution-validation subset retained six RQ1, six RQ2, and seven RQ3 groups, with no principal direction reversed. We rated confidence using study limitations, coherence, the amount and relevance of supporting studies, directness, and reproducibility or model stability. Each finding remains at moderate confidence because its direction was stable despite heterogeneous study designs, reported execution checks, and model versions.

\subsection{Overview of included publications}
\label{sec:overview}

\begin{figure}[t]
    \centering
    \begin{minipage}{0.49\textwidth}
        \centering
        \includegraphics[width=\linewidth]{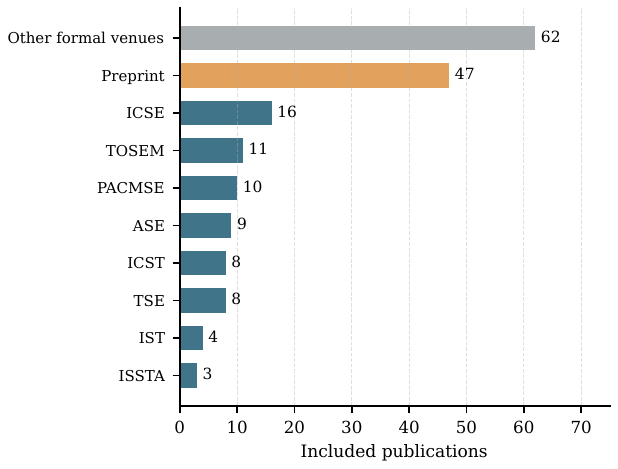}
        \captionof{figure}{Included publications by canonical venue.}
        \label{fig:dis_venues}
    \end{minipage}
    \hfill
    \begin{minipage}{0.49\textwidth}
        \centering
        \includegraphics[width=\linewidth]{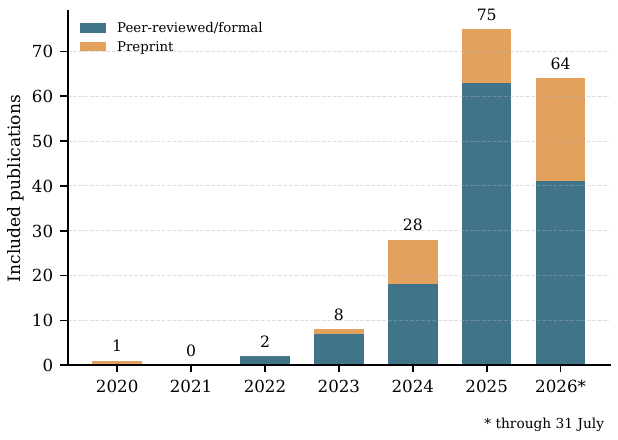}
        \captionof{figure}{Publication year by publication status.}
        \label{fig:dis_years}
    \end{minipage}
\vspace{-0.3cm}
\end{figure}

We verified each included publication's topic, analytical layer, status at the cutoff, and canonical outlet. Fig.~\ref{fig:dis_venues} reports outlet counts for the 178 included publications. Forty-seven publications remained preprints at the cutoff. Among formal outlets, ICSE contributed 16 publications, followed by the TOSEM series with 11 and PACMSE with 10. ASE contributed nine publications, ICST and TSE eight each, IST four, and ISSTA three. Other formal venues contributed 62 publications.

Regarding the publication distribution over time, Fig.~\ref{fig:dis_years} counts formal publications by the year in which the represented formal version first became available online. Here, preprints are eligible public manuscripts without a formal record at the cutoff: arXiv records use the v1 submission year, and others use their first verifiable public year. The corpus contains one publication from 2020, none from 2021, two from 2022, eight from 2023, 28 from 2024, 75 from 2025, and 64 from the partial 2026 observation window through 31 July. The count rose from eight in 2023 to 75 in 2025, but the partial 2026 window does not support extrapolation.

\subsection{Unified taxonomy and research questions}

\begin{figure}[t]
    \centering
    \includegraphics[width=\linewidth]{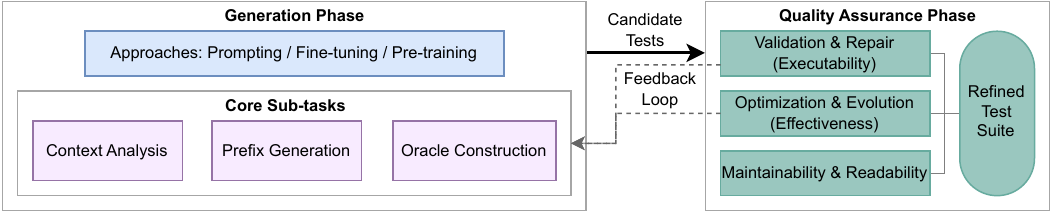}
    \caption{Lifecycle-oriented framework used to organize LLM-based unit-test generation. Colors distinguish utilization approaches, construction subtasks, and quality-assurance categories; dashed lines denote feedback. Test construction (RQ1) covers context, prefixes and inputs, and oracles, while quality assurance (RQ2) covers validation and repair, optimization and evolution, and maintainability and readability.}
    \label{fig:framework}
\vspace{-0.3cm}
\end{figure}

To analyze the collected literature, we use a lifecycle-oriented taxonomy. As illustrated in Fig.~\ref{fig:framework}, it distinguishes a generation phase that produces test artifacts from a quality assurance and enhancement phase that validates and refines them. The phases are analytical categories rather than a strictly linear sequence because many systems return execution feedback to generation.

The generation category covers methods that translate source code and project information into candidate test artifacts~\cite{schafer2023empirical,yuan2024evaluating}. It includes context analysis, prefix and input generation, and oracle construction. For context analysis, these methods use static or dynamic analysis to supply project information and reduce missing-context or compilation failures~\cite{ryan2024code,yang2024enhancing}. Prompting, fine-tuning~\cite{shang2025large,shin2024domain}, and pre-training~\cite{rao2023cat,alagarsamy2024a3test} may support one or more of these tasks. The quality-assurance and enhancement category covers mechanisms applied after or during generation. These methods use compiler or runtime feedback for repair and combine LLMs with established techniques such as search-based software testing~\cite{yuan2024evaluating,ni2024casmodatest,gu2024improving,lemieux2023codamosa,tang2024chatgpt}. These mechanisms aim to improve executability, coverage, and fault detection.

The three RQs guide comparisons within and across these categories. For RQ1, Section~\ref{sec:rq1} compares how model adaptation and project information support test construction, covering context analysis, prefix and input generation, and oracle construction. It considers each method's inputs, generated artifacts, and experimental conditions. For RQ2, Section~\ref{sec:rq2} compares validation, repair, filtering, and optimization mechanisms, including combinations with established testing techniques. It examines executability, coverage, oracle correctness, fault detection, and maintainability. RQ3 addresses limitations and research opportunities. Section~\ref{sec:rq3} examines artifact correctness, project complexity, and evaluation and deployment constraints across the taxonomy, relating them to specifications, project context and tools, hybrid techniques, and agentic workflows.

\section{Generation Methodologies}
\label{sec:rq1}

\begin{figure}[t]
    \centering
    \includegraphics[width=\linewidth]{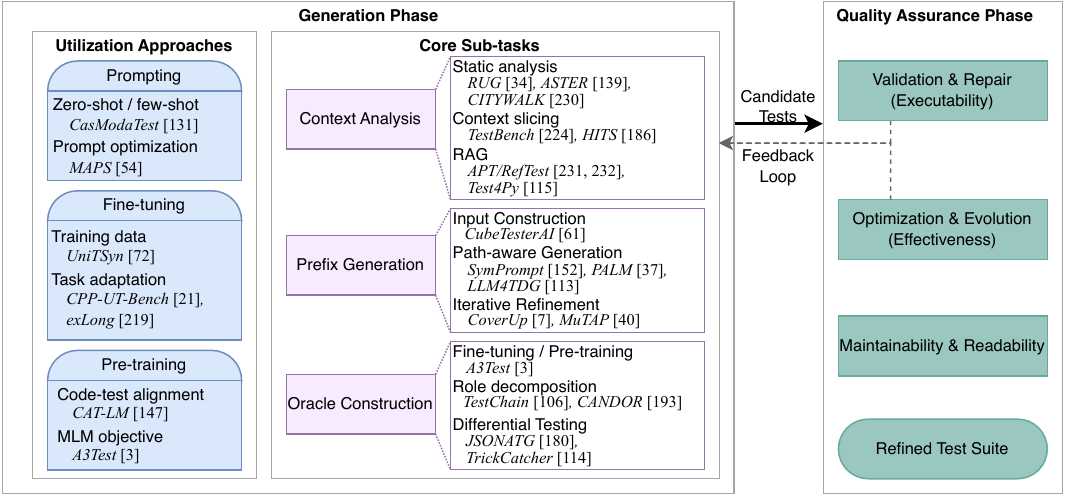}
    \caption{RQ1 framework for test construction. Prompting, fine-tuning, and pre-training can support context analysis, prefix and input generation, and oracle construction; representative systems illustrate each role.}
    \label{fig:framework_rq1}
\vspace{-0.3cm}
\end{figure}

Fig.~\ref{fig:framework_rq1} expands the generation phase shown in Fig.~\ref{fig:framework}. We first compare model-utilization approaches, then examine context analysis, prefix and input generation, and oracle construction. Of the 178 publications, 164 (92.1\%) used or centrally evaluated prompting, 48 (27.0\%) used or centrally evaluated fine-tuning or other task-specific parameter adaptation, and 5 (2.8\%) implemented pre-training or continued pre-training. These categories overlap because a publication can use several approaches. Table~\ref{tab:strategy_comparison} compares their data, computation, and deployment requirements.

Table~\ref{tab:technical_comparison} compares input requirements and techniques in fourteen workflows from this section and Section~\ref{sec:rq2}. Existing tests can provide demonstrations or an initial suite. CasModaTest uses prepared code--test examples, while TestTailor starts from a suite that another generator can produce~\cite{ni2024casmodatest,zhou2026testtailor}. CoverUp can start from an empty suite or extend existing tests~\cite{altmayer2025coverup}. CITYWALK and CoverUp both combine analysis and retrieval, but CITYWALK uses compiler errors to repair tests, whereas CoverUp uses execution errors and coverage feedback for repair and further generation~\cite{zhang2025citywalk,altmayer2025coverup}. Execution and coverage or mutation feedback are marked only when used during generation.

\subsection{Core utilization approaches}
\label{sec:strategies}

\subsubsection{Prompt engineering}
% Audited corpus synthesis: RCG-02
Prompt engineering adapts a frozen model through instructions, examples, and task context. Prompts encode scenarios, testing intentions, self-guidance, structural patterns, or tracked objectives~\cite{yin2024you,yang2025diffutester,qi2025project,nan2025intention,chowdhury2026sparc,wang2026tatg}. Retrieval supplies repository context without parameter updates, although repeated retrieval, inference, and validation still incur costs.

\paragraph{Zero-shot vs. Few-shot learning}
% Audited corpus synthesis: RCG-01
Prompt-based generators differ in instructions, demonstration selection, and task design~\cite{guilherme2023initial,li2023prompting,ni2024casmodatest,alagarsamy2025enhancing,yuan2024evaluating,andrzejewski2025automated,gao2025prompt,ouedraogo2026prompt}. Zero-shot prompting requires no project demonstrations, whereas few-shot prompting supplies examples of the desired structure or task~\cite{ni2024casmodatest,yuan2024evaluating}. JSONATG uses historically failure-revealing tests~\cite{zhong2024advancing}. CasModaTest uses separate prompts and demonstration pools for prefixes and oracles. It clusters each pool by semantic similarity and selects the example most similar to the focal method from each cluster~\cite{ni2024casmodatest}. The generated prefix becomes part of the oracle prompt. For comparison, CLAST refines existing tests into clearer in-context examples and reports improvements for RAGGen and TELPA~\cite{yang2025clarifying}.

\paragraph{Automated prompt optimization}
On 33 programs, Guilherme and Vincenzi~\cite{guilherme2023initial} compared a basic prompt with one specifying boundary-value analysis and timeout constraints. Across the tested temperatures, the best execution success rate rose from 52.5\% to 64.6\%, or 12.1 percentage points. Automated prompt optimization (APO) seeks to reduce the manual trial and error involved in such design. MAPS ranks prompt candidates by line and branch coverage on a development set and generates alternatives using different modification strategies. It groups similar failures, derives concise rules, and checks whether adding each rule improves the prompt. The final prompt combines the selected instruction, retained rules, and project context~\cite{gao2025prompt}.

\subsubsection{Model fine-tuning}
Fine-tuning methods differ in adaptation data and the proportion of model parameters updated.

\paragraph{Dataset construction and quality}
% Audited corpus synthesis: RCG-07
Fine-tuning studies construct code--test corpora for adaptation to particular languages or domains, paying attention to structural alignment and data cleaning~\cite{tufano2020unit,rao2023cat,shin2024domain,he2024unitsyn,qiu2026boosting,pipalani2025bench,zhang2025less}. Methods2Test contains hundreds of thousands of mined pairs~\cite{tufano2020unit}. UniTSyn uses the language server protocol to pair functions with their tests, producing 2.7 million samples across five languages~\cite{he2024unitsyn}. Mining can introduce syntax errors or irrelevant pairs. CleanTest found noise in about 44\% of Methods2Test samples and reported better results and lower training cost after filtering to a smaller dataset~\cite{zhang2025less}. Data curation is therefore part of the adaptation method, with consequences for test quality and training cost.

\begin{table*}[t]
\centering
\caption{Comparison of model-utilization approaches for LLM-based unit-test generation.}
\label{tab:strategy_comparison}
\resizebox{\textwidth}{!}{%
\begin{tabularx}{1.47\textwidth}{@{}>{\raggedright\arraybackslash}p{3.66cm}>{\raggedright\arraybackslash}X>{\raggedright\arraybackslash}X>{\raggedright\arraybackslash}X@{}}
\toprule
\textbf{Aspect} & \textbf{Prompting (92\%)} & \textbf{Fine-tuning (27\%)} & \textbf{Pre-training (3\%)} \\
\midrule
Core mechanism & Condition a frozen model with instructions, context, and optional examples. & Update some or all model parameters. & Train from scratch or continue pre-training on broad corpora. \\
\midrule
Training materials & Not required & Task- or domain-specific examples & Code, text, and code--test corpora \\
\addlinespace[3pt]
Training data organization & Not applicable & Paired inputs and targets, or rewards from reinforcement learning & Paired or unpaired files and method--assertion pairs \\
\midrule
Main costs & Context preparation, retrieval, and repeated model calls & Data curation and task-specific parameter training & Corpus preparation and large-scale model training \\
\midrule
Typical use & Rapid adaptation and tool integration & Language, framework, or task specialization & Model development and code--test alignment \\
\midrule
Main constraints & Context windows and missing project information & Data quality and overfitting to projects or domains & Corpus scale and computing resources \\
\bottomrule
\end{tabularx}%
}
\vspace{-0.3cm}
\end{table*}

\paragraph{Training strategies and domain adaptation}
% Audited corpus synthesis: RCG-08
Full fine-tuning (FFT) updates all model parameters, which can be computationally expensive for large models. Parameter-efficient fine-tuning (PEFT) reduces adaptation cost by updating a small set of trainable parameters. Low-rank adaptation (LoRA) keeps the original weights frozen and learns updates represented by two low-rank matrices. It has been used to generate tests for complex C++ projects~\cite{bhargava2024cpp} and exceptional behavior~\cite{zhang2024exlong}. Storhaug et al.~\cite{storhaug2024parameter} reported results comparable with FFT at lower cost in their setting. Beyond supervised learning, RLSQM uses reinforcement learning (RL) with rewards from static analysis. It improved syntactic correctness, calls to the focal method, and the absence of selected test smells over its supervised baseline~\cite{steenhoek2025reinforcement}.

Domain adaptation addresses project conventions and APIs. On SF110, Chen et al.'s multi-task prompt-learning framework achieved higher compilation success than DeepSeek-Coder~\cite{chen2025efficient}. In C\# and C++ game-development tasks, GameUnitGen-7B and GameUnitGen-13B, adapted to game-engine APIs and project code, achieved higher execution success than CodeLlama-70B~\cite{paduraru2024llm}. Ahmad and Naeem fine-tuned DeepSeek-Coder-33B on an industrial product's code--test pairs~\cite{ahmad2026boosting}. For generation from requirements, Alagarsamy et al.\ found that combining fine-tuning and their prompt design outperformed either component alone~\cite{alagarsamy2025enhancing}.

\subsubsection{Pre-training}
Pre-training and continued pre-training update parameters using large corpora and are the most resource-intensive adaptation strategies considered here. Their main focus is code--test alignment. Tufano et al.\ pre-trained a BART-based model on English and code, reducing validation loss by 25\% relative to training from scratch~\cite{tufano2020unit}. CAT-LM trained a 2.7-billion-parameter model on 1.1 million code--test file pairs and unpaired code and test files, separating paired files with a \texttt{<|codetestpair|>} token~\cite{rao2023cat}. A3Test instead uses masked-language pre-training on method--assertion pairs before fine-tuning for complete unit tests~\cite{alagarsamy2024a3test}. Thus, CAT-LM learns file-level relationships, while A3Test transfers assertion knowledge to complete tests.

\begin{table*}[t]
\centering
\caption{Input requirements and techniques used in selected LLM-based unit test generation workflows.}
\label{tab:technical_comparison}
\label{tab:technical_comparison_data}
\begingroup
\setlength{\tabcolsep}{4pt}
\resizebox{\textwidth}{!}{%
\begin{tabularx}{1.34\textwidth}{@{}>{\raggedright\arraybackslash}p{.21\textwidth}>{\raggedright\arraybackslash}p{.23\textwidth}*{6}{>{\centering\arraybackslash}X}@{}}
\toprule
& \multicolumn{2}{c}{\textbf{Input requirements}} & \multicolumn{5}{c}{\textbf{Techniques used}} \\
\cmidrule(lr){2-3}\cmidrule(l){4-8}
\textbf{Method} & \textbf{Program context} & \textbf{Existing tests} & \textbf{Program analysis} & \textbf{Context retrieval} & \textbf{Test execution} & \textbf{LLM repair} & \textbf{Coverage / mutation feedback} \\
\midrule
CasModaTest~\cite{ni2024casmodatest} & Signatures\newline Class context & $\checkmark$ & $\times$ & $\checkmark$ & $\checkmark$ & $\checkmark$ & $\times$ \\
\addlinespace[3pt]
SymPrompt~\cite{ryan2024code} & Focal source file & $\times$ & $\checkmark$ & $\times$ & $\times$ & $\times$ & $\times$ \\
\addlinespace[3pt]
RefTest~\cite{zhang2026reference} & Project code & $\bigcirc$ & $\checkmark$ & $\checkmark$ & NR & $\checkmark$ & $\times$ \\
\addlinespace[3pt]
TELPA~\cite{yang2024enhancing} & Module code & $\times$ & $\checkmark$ & $\checkmark$ & $\checkmark$ & NR & $\checkmark$ \\
\addlinespace[3pt]
LSPRAG~\cite{go2025lsprag} & Project code & $\checkmark$ & $\checkmark$ & $\checkmark$ & $\times$ & $\checkmark$ & $\times$ \\
\addlinespace[3pt]
Test4Py~\cite{liu2025llm} & Project code & $\times$ & $\checkmark$ & $\checkmark$ & $\checkmark$ & $\checkmark$ & $\checkmark$ \\
\addlinespace[3pt]
CITYWALK~\cite{zhang2025citywalk} & Code / build files\newline Documentation & $\times$ & $\checkmark$ & $\checkmark$ & $\times$ & $\checkmark$ & $\times$ \\
\addlinespace[3pt]
CoverUp~\cite{altmayer2025coverup} & Project code & $\bigcirc$ & $\checkmark$ & $\checkmark$ & $\checkmark$ & $\checkmark$ & $\checkmark$ \\
\addlinespace[3pt]
PANTA~\cite{gu2025llm} & Source code\newline Dependencies & $\bigcirc$ & $\checkmark$ & $\times$ & $\checkmark$ & $\checkmark$ & $\checkmark$ \\
\addlinespace[3pt]
TestTailor~\cite{zhou2026testtailor} & Source code\newline Dependencies & $\checkmark$ & $\checkmark$ & $\checkmark$ & $\checkmark$ & $\checkmark$ & $\checkmark$ \\
\addlinespace[3pt]
CodaMosa~\cite{lemieux2023codamosa} & Module code & $\times$ & $\checkmark$ & $\times$ & $\checkmark$ & $\times$ & $\checkmark$ \\
\addlinespace[3pt]
EvoGPT~\cite{broide2025evogpt} & Focal class code & $\times$ & $\times$ & $\times$ & $\checkmark$ & $\checkmark$ & $\checkmark$ \\
\addlinespace[3pt]
MuTAP~\cite{dakhel2024effective} & Target source code & $\bigcirc$ & $\times$ & $\times$ & $\checkmark$ & $\checkmark$ & $\checkmark$ \\
\addlinespace[3pt]
CANDOR~\cite{xu2025hallucination} & Source code\newline Descriptions & $\bigcirc$ & $\times$ & $\times$ & $\checkmark$ & $\checkmark$ & $\checkmark$ \\
\addlinespace[3pt]
\multicolumn{8}{@{}l}{Existing tests: $\checkmark$ required, $\bigcirc$ optional, $\times$ not required.} \\
\multicolumn{8}{@{}l}{Techniques: $\checkmark$ used, $\times$ not used in the described workflow, NR not reported.} \\
\bottomrule
\end{tabularx}%
}
\endgroup
\vspace{-0.3cm}
\end{table*}

\subsection{Context analysis and understanding}
\label{sec:context}

These utilization approaches often need project information beyond the focal method, including dependencies, imports, and API usage, to construct executable tests. Table~\ref{tab:context_techniques} compares how static analysis, slicing, and retrieval supply this context.

\paragraph{Static analysis for dependency resolution}
% Audited corpus synthesis: RCG-03, RCG-11
Unresolved symbols were the most common type of compilation error in TestBench and Yuan et al.'s evaluations~\cite{zhang2024testbench,yuan2024evaluating}. Adding focal-class constructors, methods, and fields reduced validation loss by more than 11\% in Tufano et al.'s study~\cite{tufano2020unit}. However, 42.41\% of generated tests had build errors on Defects4J. The authors identified missing information about other classes as a limitation and proposed project-level context. Removing information about other focal-class methods reduced compilation success across all models in Yang et al.'s ablation~\cite{yang2024evaluation}. To supply this information, repository-aware generators use static analysis, context injection, and language services to expose types, dependencies, build information, and APIs~\cite{yang2024enhancing,pan2025aster,zhang2024exlong,gorla2025cubetesterai,yin2025enhancing,lenke2026influence,li2025knowledge,yang2025reflective}. Program analysis also supplies call chains, compilable prefixes, and mocking constraints, sometimes checking intermediate constructions before assembly~\cite{zhang2025citywalk,yang2025intelliunitgen,liu2025llm,zhou2025seed,wang2026call,liu2026hieratest,xu2026based,tran2026program}. ASTER identifies mocking requirements and pre-fills test skeletons~\cite{pan2025aster}, CITYWALK parses build information such as CMake files~\cite{zhang2025citywalk}, and RUG follows Rust type dependencies, checking intermediate constructions before assembling a function-level test~\cite{cheng2025rug}. CoverUp and Siddiq et al.\ supply required imports~\cite{altmayer2025coverup,siddiq2024using}, while Yin et al.\ obtain project information from \texttt{gopls} through the language server protocol~\cite{yin2025enhancing}. Type and build information helps resolve operational dependencies, while project examples reveal usage patterns.

\paragraph{Context reduction and slicing}
% Audited corpus synthesis: RCG-04
Context selection must also account for input size and code complexity. Methods guide generation with execution paths or slices~\cite{ryan2024code,wang2024hits}, or select context within a token budget~\cite{wang2026budget}. TestBench simplifies class context through abstract syntax tree analysis, removing method bodies and variable initializations while retaining declarations~\cite{zhang2024testbench}. With CodeLlama, compilation success was 29.91\% with simple context and 23.06\% with full context, a difference of 6.85 percentage points. HITS decomposes long methods into logical slices to focus generation on selected behaviors~\cite{wang2024hits}. Selection must balance missing dependencies against the noise and cost of irrelevant material, taking account of the model budget, project structure, and target behavior.

\paragraph{Retrieval-augmented generation (RAG)}
% Audited corpus synthesis: RCG-05
RAG grounds generation in retrieved material~\cite{gao2023retrieval}. For test generation, retrieval units include project tests, types, documentation, language-server output, and structurally similar examples, with different context budgets~\cite{zhang2024llm,shin2024retrieval,xu2025clover,go2025lsprag,chu2025palm,fingleton2026enhancing,zhang2026reference}. APT, later reported as RefTest, selects project tests through structural similarity and method dependencies to illustrate local testing conventions~\cite{zhang2024llm,zhang2026reference}. Test4Py retrieves candidate Python types using parameter behavior and supplies constructors, module paths, and class summaries~\cite{liu2025llm}. For five Python ML/DL libraries, Shin et al.\ retrieve API documentation, GitHub issues, and StackOverflow questions and answers, limiting document length before prompting~\cite{shin2024retrieval}. Coverage improved without corresponding gains in correctness.

% Diagnostic content-density trial, not an approved replacement.
\begin{table*}[t]
\centering
\caption{Context-analysis techniques used to supply project information for test generation.}
\label{tab:context_techniques}
\resizebox{\textwidth}{!}{%
\begin{tabularx}{1.35\textwidth}{@{}>{\raggedright\arraybackslash}p{0.18\textwidth}>{\raggedright\arraybackslash}p{0.26\textwidth}>{\raggedright\arraybackslash}p{0.39\textwidth}>{\raggedright\arraybackslash}X@{}}
\toprule
\textbf{Type} & \textbf{Work} & \textbf{Context provided} & \textbf{Core mechanism} \\
\midrule
\textbf{Static analysis} & RUG~\cite{cheng2025rug} & Types and dependencies & Construct context in dependency order and compile intermediate code \\
\addlinespace[2pt]
 & ASTER~\cite{pan2025aster} & Types and mocking dependencies & Identify mocking requirements and pre-fill a test skeleton \\
\addlinespace[2pt]
 & CITYWALK~\cite{zhang2025citywalk} & Build configuration & Parse project build files \\
\midrule
\textbf{Context slicing} & TestBench~\cite{zhang2024testbench} & Class declarations & Remove method bodies and variable initializations \\
\addlinespace[2pt]
 & HITS~\cite{wang2024hits} & Method slices & Decompose long methods \\
\midrule
\textbf{Retrieval} & APT/RefTest~\cite{zhang2024llm,zhang2026reference} & Project tests & Select tests by structural similarity and method dependencies \\
\addlinespace[2pt]
 & Test4Py~\cite{liu2025llm} & Types and construction information & Select types by parameter behavior \\
\addlinespace[2pt]
 & Shin et al.~\cite{shin2024retrieval} & API documentation, issues, and Q\&A & Retrieve and limit document length \\
\bottomrule
\end{tabularx}%
}
\vspace{-0.4cm}
\end{table*}

\subsection{Test prefix and input generation}
\label{sec:prefix}

Test prefixes initialize objects, supply inputs, and invoke the focal method. Domain-specific input constraints and precise initialization sequences make their construction difficult.

\paragraph{Semantic input construction}
% Audited corpus synthesis: RCG-10, RCG-34
Methods for generating inputs construct concrete values, reusable fixtures, object builders, mocks, or constraints for later generation. Bug reports, input partitions, schemas, types, and mocking metadata guide these artifacts~\cite{ouedraogo2025enriching,zhou2024based,li2025based,wang2024testeval,wang2026fixturize}. Equivalence partitions and boundary values guide input design~\cite{rodriguez2026evaluating}. For fault-directed generation, Yin et al.~\cite{yin2024you} adjust attention to defective statements. LISP generates executable drivers for library APIs from input-space partitions~\cite{li2025based}. Fixturize classifies whether a focal function or method requires fixtures~\cite{wang2026fixturize}. It routes functions needing no setup directly to test generation. For the remaining functions, it refines invocation examples using runtime feedback, then uses these examples to generate tests with setup, mocks, and inputs. Its asynchronous network example shows why this preparation matters. An incomplete mock causes a TypeError because it omits the methods needed to enter and exit the asynchronous context. For tests that require external objects or mocked dependencies, ASTER identifies the relevant fields and types and incorporates this information into a test skeleton~\cite{pan2025aster}. CubeTesterAI provides focal-method, dependency, entity, and DTO information in prompts with instructions for Mockito setup~\cite{gorla2025cubetesterai}. Valid inputs alone do not ensure that a complete test compiles, exercises the intended behavior, or contains a correct oracle~\cite{yi2023exploring,lops2025assessment}. Once a complete test has been generated, model chaining can pass it to another model for refinement~\cite{kumar2025powered}.

\paragraph{Path-aware and constraint-guided generation}
% Audited corpus synthesis: RCG-12
Path-sensitive approaches expose branch conditions, semantic constraints, program graphs, symbolic results, or explicit exploration plans to focus generation on hard-to-reach behaviors~\cite{xiong2024program,xue2025distinct,liu2025llm4tdg,liao2025navigating,wu2026palm,zeng2026logic,zhang2026llmtest,amayuelas2026planning}. SymPrompt enumerates a minimized set of basis paths and constructs a separate prompt for each, using path conditions, return behavior, and calling context. With CodeGen2, it produced approximately five times as many passing tests that correctly called the focal method as its baseline completion prompt~\cite{ryan2024code}. PALM derives condition chains and project context from Rust MIR/HIR and control-flow analysis, repairing path-specific tests with compiler feedback~\cite{chu2025palm}. LLM4TDG uses constraint-dependency graphs and backtracking to handle input requirements that must be satisfied together~\cite{liu2025llm4tdg}. Structural analysis gives generation explicit goals.

\paragraph{Feedback-guided generation}
% Audited corpus synthesis: RCG-06
Coverage reports can be used to guide further generation toward code that existing tests have not exercised~\cite{altmayer2025coverup,yang2026cgtest}. To focus on the remaining targets, existing methods remove already covered code from the generation context~\cite{xu2026enhancing} or use existing tests whose execution paths are close to an uncovered target~\cite{zhou2026testtailor}. PANTA uses dynamic coverage to select among statically derived paths. Its exploitation--exploration strategy selects the path with the most uncovered statements and branches and the least frequently selected path~\cite{gu2025llm}. For fault detection, other work uses synthetic bugs derived from historical defects to guide further test generation~\cite{wang2026knowledge}, while MuTAP targets surviving mutants~\cite{dakhel2024effective}. Mokav instead compares internal variable values across two program versions, even when an input produces identical outputs. It adds values observed in only one version to the next prompt and repeats generation until an input produces different outputs or the iteration limit is reached~\cite{etemadi2024mokav}.

\subsection{Test oracle generation}
\label{sec:oracle}

An oracle checks whether the observed program behavior matches the expected behavior. The approaches discussed here are used to learn assertion patterns, reason about expected outcomes, or compare the results of different implementations. Table~\ref{tab:oracle_strategies} summarizes representative methods.

\begin{table*}[t]
\centering
\caption{Representative techniques for test-oracle construction and pseudo-oracles.}
\label{tab:oracle_strategies}
\resizebox{\textwidth}{!}{%
\begin{tabularx}{1.30\textwidth}{@{}>{\raggedright\arraybackslash}p{0.26\textwidth}>{\raggedright\arraybackslash}p{0.51\textwidth}>{\raggedright\arraybackslash}X@{}}
\toprule
\textbf{Representative works} & \textbf{Core mechanism} & \textbf{Key findings} \\
\midrule
\multicolumn{3}{@{}l}{\textit{Fine-tuning \& Pre-training}} \\
\addlinespace[2pt]
A3Test~\cite{alagarsamy2024a3test} & Pre-train on ATLAS method--assertion pairs, then fine-tune on complete tests. & Assertion knowledge transfers to complete unit tests. \\
\addlinespace[3pt]
Shang et al.~\cite{shang2025large} & Fine-tune models to generate assertions from focal methods and test prefixes. & Assertion matching does not establish oracle correctness. \\
\addlinespace[3pt]
Watson et al.~\cite{watson2020learning} & Predict assertions from test methods and focal methods. & Assertion patterns can be learned from paired code and tests. \\
\addlinespace[6pt]
\multicolumn{3}{@{}l}{\textit{Reasoning and Role Decomposition}} \\
\addlinespace[2pt]
TestChain~\cite{li2024large} & Separate input design from calculating expected outputs with a Python interpreter. & Separate calculation lets input design focus on scenarios. \\
\addlinespace[3pt]
CANDOR~\cite{xu2025hallucination} & Agents generate, execute, and repair tests and review oracles against specifications. & Agent discussion improves oracle correctness. \\
\addlinespace[6pt]
\multicolumn{3}{@{}l}{\textit{Differential Testing}} \\
\addlinespace[2pt]
JSONATG~\cite{zhong2024advancing} & Execute generated tests on fastjson2 and fastjson. & Differential checks help separate defects from test errors. \\
\addlinespace[3pt]
TrickCatcher~\cite{liu2025llmtrick} & Compare the target with generated variants on 151 EvalPlus functions. & Comparing program variants reduces the need to predict expected outputs directly. \\
\bottomrule
\end{tabularx}%
}
\vspace{-0.3cm}
\end{table*}

\paragraph{Assertion inference via fine-tuning and pre-training}
% Audited corpus synthesis: RCG-13, RCG-14
The reviewed assertion-generation literature ranges from neural and summarization-guided predictors to fine-tuned code language models operating on focal-method context~\cite{tufano2022accurate,dinella2022toga,hossain2023neural,zhang2025saga,he2024focal,wang2024chat,primbs2025assert5}. Watson et al.\ trained a sequence-to-sequence model to predict assertions from test methods and focal methods~\cite{watson2020learning}. TOGLL fine-tunes code models for assertion and exception oracles using test prefixes and focal-method information~\cite{hossain2025togll}. Other oracle models combine retrieval, joint training, and the prediction of multiple assertions to move beyond a single expected-value statement~\cite{wang2024deep,zhang2025exploring,zhang2025improving,zhang2025improvinga,li2025retrieval}. A3Test transfers assertion knowledge from ATLAS method--assertion pairs to complete test generation~\cite{alagarsamy2024a3test}. TECO trains CodeT5 to predict the next test statement from code semantics and preceding execution context, then reranks completions through execution~\cite{nie2023learning}. Fine-tuned CodeLlama-7B achieved 71.42\% exact match on Data\textsubscript{old}~\cite{shang2025large} and 72.50\% accuracy on ATLAS~\cite{zhang2025exploring}. These results support specialization, but matching a reference assertion does not establish oracle correctness.

\paragraph{Reasoning, role decomposition, and validation}
% Audited corpus synthesis: RCG-16
Prompts can target specific oracle tasks, such as detecting incorrect patches~\cite{jang2026improving}, while retrieved project information supports revision of multiple assertions~\cite{liu2025rag}. AugmenTest derives assertions from documentation and developer comments without access to the implementation~\cite{khandaker2025augmentest}. In its comparison of extended prompts and RAG variants, RAG did not consistently improve on the prompt-based approach. Chain-of-thought prompting elicits intermediate analysis, but its rationale does not guarantee a correct oracle. TestChain assigns input design and the calculation of expected outputs to designer and calculator agents, using a Python interpreter for calculation~\cite{li2024large}.

Generated assertions can then be repaired using compilation and execution feedback~\cite{hayet2025chatassert}, or filtered using generated postconditions~\cite{yu2025decon} or model self-inspection~\cite{li2026autooracle}. Nexus assigns oracle generation and review to different agents and uses execution against a model-generated implementation to guide further revision~\cite{huang2025nexus}. ChatTester generates complete tests and refines them with project context and compiler feedback~\cite{yuan2024evaluating}. CANDOR assigns agents to generate, execute, and repair tests and to review oracles against specifications~\cite{xu2025hallucination}. Compilers check whether tests build, while specification-based review checks expected behavior.

\paragraph{Differential testing as a pseudo-oracle}
Differential testing supplies a pseudo-oracle when a comparison implementation or version is available. JSONATG executes tests on fastjson2 and fastjson, treating failures specific to fastjson2 as candidate bugs for manual review~\cite{zhong2024advancing}. On fastjson2, it reported 59 bugs, of which 47 were confirmed by developers and 28 had been fixed. On 151 EvalPlus functions, TrickCatcher generates an input generator and runs its inputs across the target and generated variants. It selects the most frequent output among variants that disagree with the target as the oracle~\cite{liu2025llmtrick}. Comparing outputs reduces the need to predict expected values directly, but discrepancies require triage, and conclusions depend on the comparison target's reliability.

% \begin{tcolorbox}[colback=gray!10, colframe=black, title=\textbf{RQ1 Key Insight: Context is the New Code}]
\begin{tcolorbox}[colback=gray!10, colframe=black, title=\textbf{RQ1 Key Insight}]
\textbf{Context-enriched generation:} Many project-level generators supplement the focal method with information needed to construct a test. Type and build information supports executable setup, project tests and documentation guide API use and expected behavior, and path or execution feedback directs further input generation. The design problem is to select the information needed for each construction task and make it available to the generator. Context construction is thus central to generation, alongside model adaptation.
\end{tcolorbox}

\section{Quality Assurance and Enhancement}
\label{sec:rq2}

\begin{figure}[t]
    \centering
    \includegraphics[width=\linewidth]{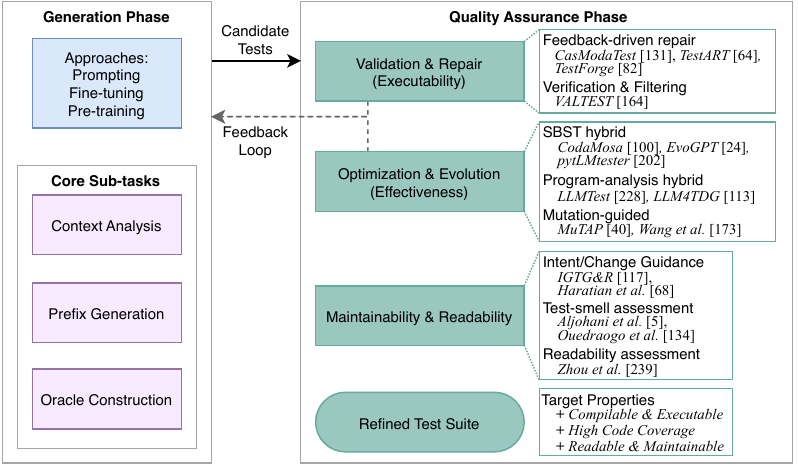}
    \caption{RQ2 framework for quality assurance. The three categories cover validation and repair, optimization and evolution, and evaluation of maintainability and readability.}
    \label{fig:framework_rq2}
\vspace{-0.3cm}
\end{figure}

Generated tests vary in executability, oracle correctness, and maintainability. Fig.~\ref{fig:framework_rq2} groups quality assurance into validation and repair, optimization and evolution, and maintainability and readability. Table~\ref{tab:qa_strategies} compares the inputs, feedback, and roles of representative methods. Work that only modifies existing tests provides context on maintenance and refactoring.

\subsection{Iterative validation and repair}
\label{sec:validation_repair}

% Diagnostic content-density trial, not an approved replacement.
\begin{table*}[t]
\centering
\caption{Representative quality-assurance techniques and their roles in test generation.}
\label{tab:qa_strategies}
\resizebox{\textwidth}{!}{%
\begin{tabularx}{1.25\textwidth}{@{}>{\raggedright\arraybackslash}p{0.22\textwidth}>{\raggedright\arraybackslash}p{0.29\textwidth}>{\raggedright\arraybackslash}p{0.31\textwidth}>{\raggedright\arraybackslash}X@{}}
\toprule
\textbf{Technique} & \textbf{Representative works} & \textbf{Inputs and feedback} & \textbf{Role in workflow} \\
\midrule
\multicolumn{4}{@{}l}{\textbf{Validation and repair}} \\
\addlinespace[2pt]
Feedback loops & CasModaTest~\cite{ni2024casmodatest}, TestART~\cite{gu2024improving}, TestForge~\cite{jain2025testforge} & Errors, execution, coverage & Repair and further generation \\
\addlinespace[2pt]
Rule repair & TestART~\cite{gu2024improving} & Recognizable errors & Templates with LLM fallback \\
\addlinespace[2pt]
Candidate filtering & VALTEST~\cite{taherkhani2024valtest} & Entropy and execution labels & Classifier filtering can improve validity but lower mutation scores \\
\addlinespace[2pt]
 & ConVerTest~\cite{taherkhani2026consistency} & Test and code candidates & Cross-execute without a trusted implementation. Higher validity with small coverage and mutation score losses \\
\midrule
\multicolumn{4}{@{}l}{\textbf{Optimization and evolution}} \\
\addlinespace[2pt]
Search-based testing & CodaMosa~\cite{lemieux2023codamosa} & Code and coverage & Generate when search stalls \\
\addlinespace[2pt]
 & EvoGPT~\cite{broide2025evogpt} & Code, errors, coverage & Seed search and add tests when search stalls \\
\addlinespace[2pt]
 & pytLMtester~\cite{yang2025llm} & Code and docstrings & Infer types and generate inputs \\
\addlinespace[2pt]
Constraint guidance & LLMTest~\cite{zhang2026llmtest}, LLM4TDG~\cite{liu2025llm4tdg} & Paths and constraints & Guide input construction \\
\addlinespace[2pt]
Mutation guidance & MuTAP~\cite{dakhel2024effective}, Wang et al.~\cite{wang2026mutation} & Surviving mutants & Generate tests for surviving mutants \\
\addlinespace[2pt]
Intent/change guidance & IGTG\&R~\cite{liu2026igtg}, Haratian et al.~\cite{haratian2026aware} & Intentions and code changes & Target intended or changed behavior \\
\midrule
\multicolumn{4}{@{}l}{\textbf{Maintainability and readability}} \\
\addlinespace[2pt]
Smell assessment & Aljohani et al.~\cite{aljohani2024fine}, Ou\'edraogo et al.~\cite{edraogo2026diffusion} & Detected test smells & Identify maintenance risks \\
\addlinespace[2pt]
Readability guidance & Zhou et al.~\cite{zhou2024based} & Parameter context & Assess and improve input values \\
\bottomrule
\end{tabularx}%
}
\vspace{-0.3cm}
\end{table*}

% Audited corpus synthesis: RCG-19
Feedback loops use compiler errors, execution failures, and coverage information to revise generated tests. They help correct references to unavailable fields, recover missing project context, and guide further generation~\cite{chen2025agenttester,guo2026diagnosing,gu2025llm,jain2025testforge,yang2025testloter,liu2026jstestcraft,bach2026library}. CasModaTest feeds compiler errors and test failures back to the model~\cite{ni2024casmodatest}. Its complete pipeline achieved 77.16\% test accuracy, defined as passing and invoking the focal method directly or indirectly. CubeTesterAI compiles and executes test methods in separate files so that one malformed test cannot block the rest~\cite{gorla2025cubetesterai}. To prepare training data, Hua et al.\ repair generated tests using errors, failures, and coverage, then compress their reasoning traces for supervised fine-tuning~\cite{hua2026synthesizing}.

% Audited corpus synthesis: RCG-24
Repair within these loops can combine model-generated revisions with deterministic rules. TestART uses templates for recognizable errors, including missing imports and assertion errors, and returns the repaired tests and coverage information to subsequent generation steps. If template repair fails, it can make an additional LLM-based repair attempt~\cite{gu2024improving}. On Defects4J, its pass rate was 78.55\%, compared with ChatUniTest's 60.05\%. TestForge instead lets an agent choose file edits, test execution, and coverage inspection when deciding how to revise a suite~\cite{jain2025testforge}. It produced at least one passing test for 84.3\% of TestGenEval programs, reported as pass@1. These workflows differ in how repair decisions are divided between fixed rules and the model.

Candidate filtering selects tests to retain after generation or repair. VALTEST derives semantic-entropy features from generated inputs and expected outputs, trains a classifier with execution-derived labels, and rejects candidates predicted to be invalid~\cite{taherkhani2024valtest}. Its threshold analysis shows that stricter filtering can improve validity but reduce mutation score by removing useful candidates. ConVerTest works without a trusted implementation by sampling completions for test stubs and cross-executing them against several generated code candidates. It groups code candidates that pass the same tests, rewarding both the number of tests passed and the number of agreeing candidates. A representative of the highest-scoring group then serves as the reference for filtering tests~\cite{taherkhani2026consistency}. Across BigCodeBench-Hard and LBPP, its verification stage raised validity by 7--39 percentage points, with losses of at most two percentage points in line coverage and mutation score. As an industrial comparison, Meta's TestGen-LLM retains additions to existing Kotlin suites only when they build, pass repeatedly, and increase coverage~\cite{alshahwan2024automated}.

\subsection{Test optimization and evolution}
\label{sec:optimization}
% Audited corpus synthesis: RCG-28, RCG-29
Once tests execute, optimization aims to improve coverage and fault detection and keep tests useful as code changes. IGTG\&R generates tests for executability and coverage, then uses an LLM to refine them through intent analysis. It infers intent from signatures and surrounding class information while omitting the focal method body, then revises or adds assertions to check that intent~\cite{liu2026igtg}. As a contextual comparison, TestGeneralizer uses an existing test and its focal method to construct a scenario template. It varies the template's settings to obtain further scenarios, generates a test for each, and refines failed tests using compiler and execution feedback~\cite{qi2026generalizing}.

\paragraph{Synergy with search-based software testing}
LLM-generated suites trailed EvoSuite in overall mutation score but outperformed it on 14 of 33 programs in Guilherme and Vincenzi's study~\cite{guilherme2023initial}. Test Wars likewise shows that comparisons depend on the property measured, examining mutation score, coverage, and reliability~\cite{abdullin2025test}. Hybrid approaches assign models different roles. CodaMosa invokes an LLM to target poorly covered functions when search stalls, using the generated tests to resume exploration. Generated code is converted into the search representation and mutated before population selection. If the population remains unchanged, CodaMosa waits for a longer stall before calling the model again~\cite{lemieux2023codamosa}. EvoGPT generates and repairs the initial population and adds tests when search stalls. It reported about 10\% higher average coverage and mutation score for passing tests than TestART and EvoSuite~\cite{broide2025evogpt}. pytLMtester uses LLMs to infer types and generate inputs in Python~\cite{yang2025llm}, while another method extracts inputs from bug reports to seed search~\cite{ouedraogo2025enriching}.

\paragraph{Integration with symbolic execution}
LLMs can assist symbolic execution by constructing solver code or by proposing concrete inputs. Wang et al.'s LLM-Sym translates constraints from Python execution paths into code that calls Z3~\cite{wang2024python}. In its LeetCode-based evaluation, generated tests reproduced 70 of 111 target traces, or 63.1\%. Jiang et al.'s separate LLMSym first generates inputs with an LLM, checks them through execution, and uses unsuccessful attempts to improve prompts and construct harnesses for constraint-based tools~\cite{jiang2024towards}. On Python PathEval, it solved 556 of 747 samples, compared with 243 for CrossHair. Wu et al.'s PALM takes another approach by enumerating paths and converting their conditions into prompts for test input generation~\cite{wu2026palm}.

\paragraph{Mutation-guided optimization}
% Audited corpus synthesis: RCG-30
Mutation-guided systems use surviving mutants or diagnostics from mutation testing to identify behaviors that the current tests do not distinguish~\cite{dakhel2024effective,foster2025mutation,wang2026mutation,straubinger2025mutation}. This feedback gives the generator a concrete target for its next test, while mutant selection determines which faults the process emphasizes. MuTAP feeds surviving mutants back to the model and reported a 93.57\% mutation score on HumanEval with synthetic mutants~\cite{dakhel2024effective}. On Refactory, its best configuration detected faults in 1,623 of 1,710 buggy human-written student submissions, compared with 1,155 for Pynguin. PRIMG uses mutant subsumption to prioritize surviving mutants for Solidity tests. Its prioritized tests killed more mutants than tests for randomly selected mutants~\cite{bouafif2025primg}. AugGPT also uses mutation scores to guide tests for student programs~\cite{zheng2025automatic}.

\paragraph{Test evolution and maintenance}
% Audited corpus synthesis: RCG-23, RCG-25, RCG-26
Production changes require updating tests or constructing new ones. Change and Cover uses production diffs or affected code to augment existing suites~\cite{zhou2026change}. Testora uses natural-language change descriptions to detect regressions~\cite{pradel2025testora}. ReAccept and TaRGET update tests using old code, a code change, and the old test~\cite{chi2025reaccept,yaraghi2025automated}. ReAccept checks compilation, execution, and coverage of changed code. Validation failures become feedback for further revisions until the test passes these checks or a preset limit is reached. It achieved 60.16\% accuracy in identifying and updating obsolete tests, compared with CEPROT's 31.62\%~\cite{chi2025reaccept}. NIODebugger addresses a different problem, using runtime observations and guided edits to repair tests that pass initially but fail on subsequent executions in the same environment~\cite{ke2025niodebugger}. Stack traces and execution logs guide the agent to relevant source code for locating shared-state pollution and resetting the affected state. These studies provide context for synthesis work such as Haratian et al.'s method, which generates new tests in response to code changes~\cite{haratian2026aware}.

\subsection{Test maintainability and readability}
\label{sec:refactoring}

Obscure names, unreadable inputs, redundant setup, and test smells impede test review and maintenance. Aljohani et al.\ and Ou\'edraogo et al.\ characterize smell prevalence and variation~\cite{aljohani2024fine,edraogo2026diffusion}. Zhou et al.\ use source-code context to assess whether primitive inputs fit their parameters' meaning and guide more readable values~\cite{zhou2024based}. These studies concern newly generated tests. Refactoring studies address readability in existing tests while checking execution and behavior preservation.

% Audited corpus synthesis: RCG-27
Magic numbers and assertion roulette can hinder understanding of generated tests~\cite{edraogo2026diffusion}. UTRefactor detects smells in existing tests and gathers context about the tested code~\cite{gao2025automated}. It combines smell definitions, rules in a domain-specific language, and CoT prompting. These rules include replacing numeric literals with named variables and adding explanatory assertion messages. For tests with multiple smells, the prompt asks the model to recheck each smell and revise again if necessary. Across six Java projects, UTRefactor reduced the number of detected smells from 2,375 to 265, an 89\% reduction. The refactored tests had average compilation and execution pass rates of 95\% and 89\%, respectively. Line and branch coverage were unchanged for tests that still passed.

Readability can also be assessed through developers' work with tests. UTGen contextualizes test data, improves names, and adds comments~\cite{deljouyi2024leveraging}. It refines input values during EvoSuite's search, before test minimization and assertion generation. The LLM then improves names and adds Given--When--Then comments explaining setup, action, and expected outcomes, with instructions to preserve the data and logic at this stage. In a controlled experiment with 32 participants, developers using UTGen tests fixed more bugs. Mean completion times were shorter, described as up to 20\% faster, but the time effect was not statistically significant ($p=0.063$). Within the 30-minute limit, the time difference was larger for JSWeaponData than Budget. Participants valued Given--When--Then comments but found some comments too general to explain the tests.

\begin{tcolorbox}[colback=gray!10, colframe=black, title=\textbf{RQ2 Key Insight}]
\textbf{Engineering feedback in quality assurance:} Many reviewed systems couple LLM generation with compilers, test runners, program-analysis or search tools, and mutation or coverage feedback. These loops can improve executability, but they do not by themselves establish semantic correctness, fault-detection effectiveness, or industrial reliability.
\end{tcolorbox}

\section{Challenges and Opportunities}
\label{sec:rq3}

\begin{figure}[t]
    \centering
    \includegraphics[width=\columnwidth]{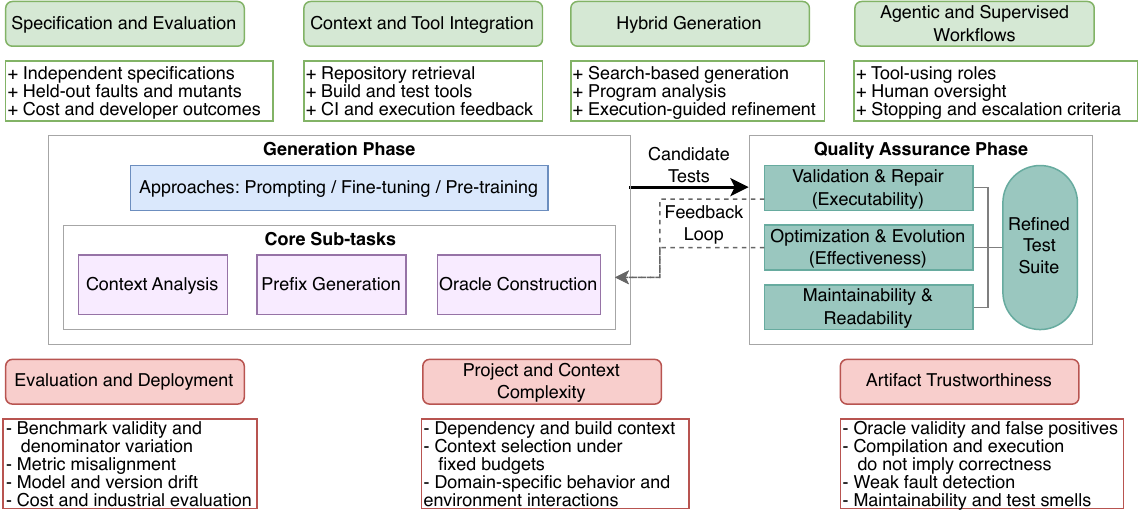}
    \caption{Challenges and research directions identified in RQ3. The figure places three challenge groups around four parallel directions concerning specification and evaluation, context and tool integration, hybrid generation, and supervised agentic workflows.}
    \label{fig:framework_rq3}
\vspace{-0.3cm}
\end{figure}

The remaining challenges concern artifact trustworthiness, CUT and project complexity, and evaluation validity. Project complexity can make it harder to construct a valid test, while incomplete evaluation can hide whether similar improvements can be achieved in other tasks or repositories. Fig.~\ref{fig:framework_rq3} relates them to parallel research directions in specifications and evaluation, project context and tools, hybrid generation, and agentic workflows.

\subsection{Challenges in automated test generation}
\label{sec:challenges}

\subsubsection{Trustworthiness and correctness of test artifacts}

Trustworthiness involves executable tests, accurate oracles, fault detection, and maintainability. A generated suite can satisfy one of these requirements while leaving others unresolved.

\paragraph{Low usability and high error rates}
In AgoneTest's initial configurations, build success ranged from 7.1\% to 36.0\% across three models and zero- or few-shot prompts. Supplying the focal class's path raised these rates by 6.2 to 28.6 percentage points~\cite{lops2025assessment}. On Defects4J, the highest test pass rate among Shang et al.'s fine-tuned models was 36.02\%, with frequent build and syntax errors~\cite{shang2025large}. Yi et al.\ also reported syntax errors and non-existent method calls, with success depending on prompts and project complexity~\cite{yi2023exploring}. Repair loops (Section~\ref{sec:validation_repair}) address these failures at the cost of further model calls, compilation, and execution~\cite{ni2024casmodatest,gu2024improving}.

\paragraph{The oracle problem and false positives}
% Audited corpus synthesis: RCG-15
Syntax checks and text matching cannot establish oracle correctness~\cite{hossain2023neural}. Work on assertion datasets, postconditions, and benchmarks with multiple program variants makes expected behavior an explicit evaluation target~\cite{shin2024assessing,liu2023more,konstantinou2024generate,molinelli2026generate,bodicoat2025understanding,prasetya2026talk,tasnim2026togbench}. Assertion errors accounted for 85.5\% of failed executions in Yuan et al.'s analysis~\cite{yuan2024evaluating}. Numerical tasks pose particular difficulties. Chang et al.\ report problems with arithmetic and loops~\cite{chang2025systematic}, while Koziolek et al.\ obtained no correct assertions for CRC\_GEN despite full statement coverage with both prompts~\cite{koziolek2024automated}. However, failing assertions are not necessarily invalid. Mathews et al.\ show that filtering tests that fail on the current implementation can discard tests exposing its defects~\cite{mathews2024design}.

Tests generated mainly from implementation code may reproduce a defect rather than expose it. Huang et al.\ compared generation from correct and incorrect code, measuring test accuracy against the datasets' correct implementations. They found that incorrect input code could lead tests to accept buggy behavior~\cite{huang2024measuring}. Documentation provides another source of intent~\cite{hossain2025doc2oracll}, but ambiguous requirements and disagreements with the implementation still require resolution.

\paragraph{Weak fault detection capability}
% Audited corpus synthesis: RCG-36
Fault detection requires distinguishing faulty behavior, not simply executing code. Reliability studies examine buggy input code, the validity of generated tests, and the ability to expose real faults~\cite{huang2024measuring,taherkhani2024valtest,chen2026context,zhao2026evaluating}. On Defects4J, the best-performing model in Shang et al.'s comparison, DeepSeek-Coder-6b, detected 8 of 163 bugs with 0.74\% precision~\cite{shang2025large}. Here, a bug-revealing test had to fail on the buggy version and pass on the fixed version. Using the reported precision, this corresponds to roughly 135 failing candidates per bug-revealing test. The low precision makes failure diagnosis part of the practical cost of generation.

\paragraph{Test smells and maintainability}
% Audited corpus synthesis: RCG-35
Test smells flag potential maintenance problems~\cite{aljohani2024fine,edraogo2026diffusion}. Magic number test denotes unexplained constants in assertions, while assertion roulette involves multiple unexplained assertions in one test method. In an HPC study, magic number test occurred in 100\% of generated tests and the human baseline~\cite{karanjai2024harnessing}, and approached 100\% in several Java configurations~\cite{ouedraogo2024large,ouedraogo2026prompt}. Another Java study found class-level assertion roulette rates of 54.77\% with TsDetect and 75.83\% with JNose under zero-shot prompting~\cite{edraogo2026diffusion}. Thus, test granularity, detectors, and human baselines matter when interpreting smell counts.

\subsubsection{Complexity of code under test (CUT)}

Project dependencies, complex control flow, and domain-specific behavior affect generation context and test execution environments. The selection methods in Section~\ref{sec:context} must work within finite context windows, while truncation can remove necessary information~\cite{tang2024chatgpt}. CAT-LM's 8,192-token window accommodated about 82\% of its Java and Python code--test file pairs~\cite{rao2023cat}. Even with complete project code in the prompt, MultiFileTest reports execution failures caused by a missing NumPy import~\cite{wang2025projecttest}. The challenge is to select and correctly use sufficient, trustworthy context within budget, not simply increase context length.

State transitions, numerical calculations, and parallel execution create different difficulties for test generation. Koziolek et al.'s SEQUENCE\_8 PLC block, containing 16 timers, reached 58.6\% statement coverage, but the environment did not simulate expiring timers~\cite{koziolek2024automated}. This limits what the result says about generation alone. Mokav reported failures on cases requiring precise mathematical calculations~\cite{etemadi2024mokav}. In HPC testing, Karanjai et al.\ found missing or incorrect OpenMP pragmas, erroneous method calls, and passing tests that did not explicitly check parallelism~\cite{karanjai2024harnessing}. Inputs, oracles, and execution environments therefore need to support the domain together.

Generation in a new project can also depend on the availability of local examples. JSONATG uses historically failure-revealing demonstrations~\cite{zhong2024advancing}, while Shin et al.\ adapt models with developer tests~\cite{shin2024domain}. Projects with little testing history may lack these examples. TestGen-LLM's authors suggested that Facebook's roughly tenfold larger pool of human-written Kotlin tests might explain its slightly higher success rate than Instagram~\cite{alshahwan2024automated}. Testing adaptation on projects excluded from training, with available examples reported, would help distinguish cross-project generalization from dependence on local testing history.

\subsubsection{Evaluation validity and ecosystem gaps}
% Audited corpus synthesis: RCG-22
Model, prompt, and version choices can change the apparent benefit of a generation technique~\cite{yang2024evaluation,ouedraogo2026prompt,konstantinou2026how}. Evaluation must therefore identify the task, baseline, and deployment conditions under which an improvement is useful.

\paragraph{Data contamination and benchmark validity}
% Audited corpus synthesis: RCG-09, RCG-31
Benchmarks now cover class-level, repository-level, and reasoning-intensive generation tasks~\cite{zhang2024testbench,jain2024testgeneval,chang2025systematic,huang2026benchmarking}. MultiFileTest includes dependencies across project files, TestGenEval uses repository test suites, and CLOVER evaluates context lengths from 4k to 128k tokens~\cite{wang2025projecttest,jain2024testgeneval,xu2025clover}. SWT-Bench and TestEval further cover bug-reproduction workflows and path-sensitive generation~\cite{mundler2024swt,wang2024testeval}. Adversarial test generation and methods targeting surviving mutants add further evaluation settings~\cite{chang2026mutant,khalili2026llmutantkiller}. Interpreting their results requires knowing which mutants or other feedback were exposed during generation and which cases were reserved for evaluation. These benchmarks broaden the tasks used for evaluation, but task complexity alone does not determine whether a model has encountered the material.

Undisclosed training data make benchmark exposure difficult to establish. Public datasets such as Defects4J may overlap with training corpora, allowing memorized artifacts to influence results~\cite{shang2025large,wang2024software}. Azanza et al. selected functions and classes from projects without publicly accessible test suites to limit reproduction of existing tests~\cite{azanza2025tracking}. Reproducible comparisons need dataset and model versions, separate development and evaluation examples, and reassessment after model updates.

\paragraph{Metric misalignment}
% Audited corpus synthesis: RCG-21, RCG-37
Different subjects and success criteria complicate comparisons among LLMs, search-based generators, symbolic execution, and prompts~\cite{schafer2023empirical,tang2024chatgpt,yuan2024evaluating,li2024large,guerino2025exploring,abdullin2025test}. Text-matching scores compare generated tests with reference tests. For complete test generation on Methods2Test$_{filter}$, Shang et al.\ reported exact-match rates no higher than 0.55\%~\cite{shang2025large}. Yet a method can admit several valid tests, so divergence from a reference need not imply failure. Conversely, textual similarity cannot determine oracle correctness or fault detection. Coverage and mutation scores also depend on the testing objective. Zhao et al.\ found them useful for regression-style comparisons but unreliable when generation used buggy code to expose its defects~\cite{zhao2026coverage}. Evaluation frameworks combine execution, cost, and other quality measures~\cite{siddiq2024using,nguyen2025evaluation,lira2025evaluating,alves2025quality,azanza2025tracking,cuze2026evaluation}. Compilation and pass rates address executability~\cite{shin2024assessing,wang2024testeval,tasnim2026togbench}. Assessing test value additionally requires measures suited to the objective, including oracle correctness, mutation or held-out fault detection, maintainability, and cost, with failed generations included.

\paragraph{High cost and practical constraints}
% Audited corpus synthesis: RCG-38
Workflow costs include adaptation, retrieval, repeated generation, build and test execution, and human review. Prompt efficiency, latency, carbon emissions, and energy measurements capture different parts~\cite{onishi2025balancing,kumari2026deceat,kumari2026sustainabilitya}. ChatTester averaged about 99 seconds per test for generation and refinement, with about 15 seconds for initial generation and 30 per refinement iteration~\cite{yuan2024evaluating}. Its authors attributed much of this time to API latency and compilation or execution. Privacy and intellectual-property requirements can constrain deployment for proprietary code. Local open-source models offer an option~\cite{zhong2024advancing}, with infrastructure and maintenance costs that must be included when comparing workflows.

\subsection{Future opportunities}
\label{sec:opportunities}

Models propose tests and interpret project information, compilers and runners expose execution failures, and specifications and reviewers help determine intended behavior. TestForge's iteration study, CodaMosa's search triggers, and work on documentation and user feedback illustrate different divisions of responsibility~\cite{jain2025testforge,lemieux2023codamosa,hossain2025doc2oracll,lahiri2022interactive}. Comparisons under matched tasks and resource budgets can identify which components contribute enough useful tests to justify their cost.

\subsubsection{Agentic testing workflows}

% Audited corpus synthesis: RCG-18, RCG-20, RCG-42
Agentic testing workflows allow a model to invoke development tools and revise tests over multiple steps. SocraTest presented a vision of conversational testing agents and an example of developer--LLM dialogue~\cite{feldt2023towards}. Agentic generators distribute planning, generation, execution, and review across specialized or tool-using agents, including repository and HPC settings~\cite{zhong2024advancing,xu2025hallucination,karanjai2025hpcagenttester,ghosh2026retry,xie2026maco,zhang2026multi}. Design choices include combining agents' tests, requesting further generation or analysis, and validating candidates before developer review~\cite{arnaudo2026black,gu2026scate,rastenis2026industry}. Order matters because generating faulty code first can lead subsequent tests to accept it~\cite{konstantinou2026risk}. Recording the inputs and outputs exchanged by agents and tools could locate such errors and trace their propagation. Explicit tool interfaces would help reproduce these interactions.

TestForge refines an initial zero-shot suite with execution and coverage feedback. On TestGenEval, coverage gains diminished after roughly ten iterations~\cite{jain2025testforge}, motivating experiments on stopping rules based on recent gains. CANDOR combines coverage-guided planning with specification-based oracle review. On its Java datasets, removing the planner reduced coverage, while removing panel discussion reduced oracle correctness~\cite{xu2025hallucination}.

% Audited corpus synthesis: RCG-32
Feedback can revise tests during a run or train models for later generation. Reinforcement learning uses static quality scores or coverage rewards~\cite{steenhoek2025reinforcement,zhang2026chain,wang2026testdecision}, while other workflows use repair templates or agreement among candidates to revise or select tests during generation~\cite{gu2024improving,khelladi2025unify,taherkhani2026consistency}. Retaining failures, successful repairs, or project conventions across runs could support later tasks. Comparisons with and without this memory could track repeated mistakes and check for misleading information after code changes to distinguish lasting adaptation from extra attempts at one test.

\subsubsection{Hybrid integration with established testing techniques}

% Audited corpus synthesis: RCG-33
Hybrid designs must decide which parts of exploration and checking to delegate to established tools, and when model-generated candidates are useful within that process. Search-based hybrids use models for stall recovery, population initialization, type inference, and mutation~\cite{lemieux2023codamosa,xiao2024optimizing,broide2025evogpt,yang2025llm,fu2026fusing}. The remaining question is whether another model call adds more coverage or reveals more faults than continued search under the same budget. Comparisons can vary trigger policies and ablate model roles, accounting for both inference and execution costs. For symbolic hybrids (Section~\ref{sec:optimization}), comparisons can contrast models that translate constraints into solver code with those that propose inputs directly~\cite{wang2024python,jiang2024towards,wu2026palm}. Relevant outcomes are whether inputs reach the intended path, how failures guide subsequent attempts, and the cost of each division of work.

\subsubsection{Specification-driven testing}
% Audited corpus synthesis: RCG-17, RCG-39
Documentation, requirements, user stories, and formalized specifications supply intent for oracle construction~\cite{jiang2024executable,molinelli2025tratto,liu2026faithful,fu2026based}. Doc2OracLL found method descriptions and \texttt{@return} tags to be the most useful Javadoc components~\cite{hossain2025doc2oracll}. Req2Test generates unit tests from user stories and acceptance criteria through structured test planning~\cite{franca2026req2test}. As a comparison, property-based systems generate properties and input strategies, considered separately from concrete-example suites here~\cite{li2026beyond}. Each generated assertion could be linked to the requirement it checks, with the relevant specification text retained alongside it. For example, an assertion expecting an exception for empty input could be linked to the requirement that specifies this behavior. This would make the basis for the expected result available during review and help diagnose disagreements between tests, code, and documentation. Incomplete, ambiguous, or outdated specifications still require checking the interpretation and executable test. When requirements change, these links could identify tests to revisit.

\subsubsection{Expansion across the testing ecosystem}

% Audited corpus synthesis: RCG-40, RCG-41
LLM applications span several testing levels and domains~\cite{wang2024software}. Jiang et al.'s Python PathEval experiments and TrickCatcher's EvalPlus experiments focus on individual functions~\cite{jiang2024towards,liu2025llmtrick}. Vertical integration across testing levels could translate failures from existing integration or system testing into focused unit tests. This requires identifying the responsible unit, reproducing its state, and constructing an oracle. Evaluation should distinguish reproduction of the original failure from coverage and fault detection of the resulting unit suite.

% Audited corpus synthesis: RCG-43, RCG-44, RCG-45
Studies of C, C++, Rust, games, and web components expose different build and context requirements~\cite{bhargava2024cpp,cheng2025rug,paduraru2024llm,chen2025multi,liu2025strut}. HPC and production projects add execution and integration demands~\cite{karanjai2024harnessing,tu2026understanding}, while educational work uses tests to identify errors in student programs~\cite{zheng2025automatic}. A practical direction is to separate reusable generation procedures from project-specific language, framework, and domain information. Tools could obtain build commands, fixture conventions, and accepted test examples from the target project, reducing workflow redesign for each codebase. Cross-project comparisons could then measure the setup and review effort that remains.

% Audited corpus synthesis: RCG-46
Human participation supplies testing intent and resolves uncertain expectations. In interactive code generation, TiCoder requests feedback on tests to guide code selection, with a user study examining code judgments and cognitive load~\cite{lahiri2022interactive}. UTGen's developer experiment examines how clearer test data, names, and comments support bug fixing~\cite{deljouyi2024leveraging}. These roles suggest requesting feedback where context or expected behavior is ambiguous and retaining answers with the tests. Comparing targeted questions with approval of every test could clarify the trade-off between developer interruptions and unresolved oracle errors.

\begin{tcolorbox}[colback=gray!10, colframe=black, title=\textbf{RQ3 Key Insight}]
\textbf{Trustworthiness and practical value:} Progress requires tests that run, check intended behavior, expose faults, and remain useful to developers. Project context, independent sources of intent, and evaluation of failures and workflow costs all contribute to these goals. Agentic and hybrid workflows offer ways to combine these components, while human review remains important where testing intent or oracle correctness is uncertain.
\end{tcolorbox}

\section{Threats to Validity}
\label{sec:threats}

\paragraph{\textbf{Threats to study selection}}
Search terms, incomplete abstracts, and gaps in database coverage may cause relevant studies to be missed. Screening judgments and multiple reports of the same work can also affect selection. To reduce these risks, we combined four core databases, two supplementary sources, venue audits, and bidirectional snowballing. Two authors screened independently and reconciled decisions, with reports linked before counting publications. Shared biases may remain, and the requirement for accessible English full texts excludes non-English and unavailable reports. Our scope also requires unit-test generation as a primary or co-primary objective and a newly constructed unit-level artifact. Standalone repair, refactoring, migration, and improvement of existing suites provide context only.

\paragraph{\textbf{Threats to data extraction and synthesis}}
Subjective interpretation may cause extraction errors or inconsistent taxonomy assignments, particularly because extraction was not independently duplicated across the full provisional corpus. The 60-publication audit checked extraction, quality, taxonomy, and execution classifications, while both authors independently classified 20 of the audited publications without seeing the existing classifications. These checks confirmed the records, but boundaries between context engineering, generation feedback, and downstream enhancement remain interpretive. Errors outside the audited sample may remain undetected.

\paragraph{\textbf{Threats to external validity}}
Concentration in Java, Python, and a few benchmarks limits conclusions about other languages, projects, and industrial settings. The 47 publications represented by preprints at the cutoff (26.4\% of 178) may also change before formal publication. No principal finding reversed in analyses of the 131 formal publications and 83 with explicit execution results. However, these subsets retain the corpus's language and benchmark limitations. Confidence remains moderate given heterogeneous settings, uneven support across subtopics, and limitations in direct validation and reproducibility.

\paragraph{\textbf{Threats from technical timeliness and model opacity}}
Model and tool updates can make results outdated, while undisclosed training data and undocumented updates hinder reproduction. To limit dependence on particular model versions, we synthesize generation methods and findings across studies. Reported performance remains tied to each model version, prompt, repository, and evaluation date. Searches on 18 August 2026 covered reports available through 31 July 2026. Later work is outside scope, and delayed indexing may have hidden eligible earlier reports.

\section{Related Work}
\label{sec:related}

Related surveys address automated test generation, LLMs across software engineering, and their use in testing and code generation. Table~\ref{tab:related_work_comparison} compares selected surveys with this review by coverage period, scope, and analytical focus.

\begin{table*}[t]
\centering
\caption{Comparison with related surveys on LLMs and automated unit testing.}
\label{tab:related_work_comparison}
\resizebox{\textwidth}{!}{%
\begin{tabularx}{1.32\textwidth}{@{}>{\raggedright\arraybackslash}p{0.29\textwidth}>{\raggedright\arraybackslash}p{0.22\textwidth}>{\raggedright\arraybackslash}p{0.30\textwidth}>{\raggedright\arraybackslash}X@{}}
\toprule
\textbf{Survey} & \textbf{Coverage period} & \textbf{Scope} & \textbf{Analytical focus} \\
\midrule
Fan et al.~\cite{fan2023large} & Through 2023 & LLMs across software engineering & Challenges and open problems, including combinations with established techniques \\
\addlinespace[4pt]
Zhang et al.~\cite{zhang2023survey} & Through 2024 & LLMs across software engineering & Code models, SE tasks, and integration into development activities \\
\addlinespace[4pt]
Chen et al.~\cite{chen2024survey} & Through 2024 & Evaluation of LLM-generated code & Metrics and benchmarks for correctness, efficiency, and readability \\
\addlinespace[4pt]
Wang et al.~\cite{wang2024software} & Through 2023 & LLMs for unit/system testing and program repair & Testing tasks, models, prompts, inputs, and accompanying techniques \\
\addlinespace[4pt]
Zapkus and Slotkien\.{e}~\cite{zapkus2024unit} & Through 2024 & LLMs for unit-test generation & Model use, drawing also on broader code-generation studies \\
\addlinespace[4pt]
Zhang et al.~\cite{zhang2025large} & Through March 2025 & LLM-based test generation, oracle generation, and test evolution & Organization by technology and task, including hybrid techniques \\
\addlinespace[4pt]
Chen et al.~\cite{chen2026survey} & 2020--May 2025 & Automated unit-test generation & Comparison of five generation paradigms \\
\midrule
\textbf{This review} & 2020--31 July 2026 & 178 publications on constructing new unit-level test artifacts & Test construction and quality assurance organized by lifecycle. Standalone maintenance, refactoring, and broader testing provide context. \\
\bottomrule
\end{tabularx}%
}
\vspace{-0.3cm}
\end{table*}

Anand et al.\ survey symbolic, model-based, combinatorial, random, and search-based test generation~\cite{anand2013orchestrated}. Surveys of unit-testing practice describe developers' difficulties in writing and maintaining tests~\cite{daka2014survey,runeson2006survey}. Amalfitano et al.\ map 20 secondary studies on established AI techniques for dynamic testing, including generation, test-data definition, optimization, and oracles~\cite{amalfitano2023artificial}. These reviews inform hybrid generation and describe practical demands on generated tests.

Among broader LLM reviews, Zhang et al.~\cite{zhang2023survey} map 926 studies to 112 software-engineering tasks through August 2024, organizing them by code models, tasks, and integration issues. Fan et al.~\cite{fan2023large} discuss challenges across the field, including combinations of LLMs with established software-engineering techniques. Within testing, Wang et al.~\cite{wang2024software} organize 102 studies through October 2023 by testing task, models, prompts, inputs, and accompanying techniques, covering unit-test generation, system testing, and program repair. Chen et al.~\cite{chen2024survey} examine metrics and benchmarks for generated code's correctness, efficiency, and readability. These benchmarks use tests to judge generated programs. We examine how tests and oracles are constructed and checked.

Three closer reviews address automated unit testing. Zapkus et al.~\cite{zapkus2024unit} discuss LLM adoption for unit-test generation, drawing also on broader code-generation work. Zhang et al.~\cite{zhang2025large} organize 105 studies through March 2025 by technology and task, including generation, oracles, evolution, and combinations with program analysis, repair, search, and mutation testing. Chen et al.~\cite{chen2026survey} compare five paradigms from 2020 through May 2025, covering search, symbolic execution, specifications, smaller pretrained models, and LLMs. Our review covers work through 31 July 2026 in which unit-test generation is a primary or co-primary objective. It follows the constructed unit-level artifact through context preparation, prefix and input generation, oracle construction, and validation and repair, relating generation inputs to revision feedback. Established techniques enter this analysis through hybrid workflows. Other work on standalone maintenance, refactoring, industrial improvement, and broader testing supplies contextual comparisons.

\section{Conclusion}
\label{sec:conclusion}

This review synthesizes 178 publications on LLM-based unit-test generation in a lifecycle framework connecting test construction with quality assurance. First, retrieval, static analysis, slicing, and feedback supply the project dependencies and conventions needed to construct usable tests. Second, generation is coupled to compilers, test runners, program analysis, and search. Compilation and execution failures guide repair, coverage and mutation feedback guide further generation, and filtering selects tests for the suite. However, repeated repair adds execution costs, and filtering may discard useful tests. Third, agentic and hybrid workflows are expanding alongside persistent problems with assertions, maintainability, benchmark validity, and the gap between coverage and fault detection. Industrial use also requires managing dependencies, build environments, and costs.

Oracle correctness remains a central challenge because assertions can agree with faulty implementations. Alongside compilation and execution checks, expected outputs should be reviewed using specifications or developer feedback. Tests against known faults reveal whether generated suites detect incorrect behavior. Comparing original and repaired tests on the same faults can show whether gains justify the additional execution and review effort.

LLMs now support unit-test generation, but their practical role depends on reproducible evaluation, independent validation of generated behavior, and appropriate human or tool supervision.

% \begin{acks}
% \end{acks}

% \nocite{*}

\bibliographystyle{ACM-Reference-Format}
\bibliography{ref}

% \appendix

\end{document}